\documentclass{article}
\usepackage{bezier,emlines2,gdgspace}
\advance\textheight by -\headsep\advance\textheight by -\headheight
\begin{document}
{\Large Shutters, Boxes, But No Paradoxes: 
Time Symmetry Puzzles in Quantum Theory} \vskip .5cm
RUTH E. KASTNER\vskip .5cm
\noindent Department of Philosophy \\
University of Maryland \\
College Park, MD 20742 USA. \\
\rule{5cm}{.5pt}
\vspace{.1cm}

Abstract. The ``N-Box Experiment'' is a much-discussed 
thought experiment in quantum mechanics. It is claimed by some authors
that a single particle prepared in a superposition of N+1 box locations and
which is subject to a final ``post-selection'' measurement 
corresponding to a different superposition can
be said to have occupied ``with certainty'' N boxes during the
intervening time. However, others have argued
that under closer inspection, this surprising claim fails to hold. 
Aharonov and Vaidman have continued their advocacy of the
claim in question by proposing a variation on the N-box experiment,
in which the boxes are replaced by shutters and the
pre- and post-selected particle is entangled with a photon.
These authors argue that the resulting ``N-shutter experiment''
strengthens their original claim regarding the N-box
experiment. It is argued in this paper that the apparently surprising
features of this variation are no more robust than those of the N-box experiment
and that it is not accurate to say that the particle is
``with certainty'' in all N shutters at any given time.

\vskip 1cm\large

	In traditional quantum mechanics, systems are
prepared in a chosen state and then subjected to measurements. One then
studies these systems using the
Born Rule for the probability of outcomes conditional
on the chosen state.\footnotemark[1] This
method is time-asymmetric, having a boundary condition
(the prepared state) only in the past. A time-symmetric way to study quantum
systems is through pre- and post-selection, in which a system
is subjected both to the usual preparation at an initial time $t_1$
and also to a final
``post-selection'' measurement at $t_2$, performed after an intervening
measurement at time $t$, $t_1 < t < t_2$. 

	Y. Aharonov and L. Vaidman have proposed a
 thought experiment that they say indicates
 the ``robustness'' of surprising features  
claimed to hold for  pre- and post-selected systems (Aharonov
and Vaidman 2002). The particular surprising feature they refer to is
their claim, applied to a well-known
gedanken experiment known as the ``N-box experiment,'' that 
``with utilization of a particular pre- and post-selection,
we can claim that [a] particle should have been
found with certainty in each one out of several places given
that it was looked for only in that place.'' (Aharonov
and Vaidman (2002, p.1).
 However, this new experiment, involving a particle having access to
 $N+1$ different shutters and a photon emitted toward $N$ of those
shutters, ceases to be surprising on closer inspection, for the same
 reasons already discussed with regard to the N-box example (cf. Cohen (1995), 
 Griffiths (1996),
Kastner (1999a,b),  replies by Vaidman (1999a,b).)

	What we have in the proposed experiment is essentially the claim that the shutter 
particle, which blocks the passage of a photon it encounters, can be thought of as occupying
 ``all N shutters at once.'' However, the assertion by Aharonov
and Vaidman that the photon ``indicates the presence of the shutter [particle] in
every slit''\footnotemark[2]
 is misleading in the sense that the shutter particle
cannot be said to be ``with certainty'' in all the shutters; all that
can be concluded is that the shutter particle and the photon are
in the same superposition of shutters.

	To see this more clearly, let $N=2$
 and look at this simple variant, which 
tradition calls the ``3-shutter experiment.''\footnotemark[3]
Figure 1 shows a diagram of the
Hilbert space of the shutter particle, whose pre- and post-selected states $|\psi_1\rangle$
and $|\psi_2\rangle$ are the
same as in the three-box example. 

\special{em:linewidth 0.1pt}
\unitlength 1.00mm
\linethickness{0.1pt}
\begin{picture}(106.33,131.00)(0,15)
\unitlength 1.00mm
\linethickness{0.1pt}
\emline{56.33}{103.67}{1}{85.00}{103.67}{2}
\emline{85.00}{103.67}{3}{85.00}{46.67}{4}
\emline{85.00}{46.67}{5}{56.67}{46.67}{6}
\emline{56.67}{46.67}{7}{56.67}{76.33}{8}
\emline{64.00}{96.33}{9}{35.00}{96.33}{10}
\emline{35.00}{96.33}{11}{56.33}{103.33}{12}
\emline{63.33}{96.33}{13}{84.67}{103.00}{14}
\emline{64.00}{38.67}{15}{85.33}{46.67}{16}
\emline{63.67}{96.33}{17}{63.67}{38.67}{18}
\emline{63.67}{38.67}{19}{34.67}{38.67}{20}
\emline{34.67}{38.67}{21}{34.67}{96.67}{22}
\emline{35.00}{38.67}{23}{56.33}{46.67}{24}
\emline{35.00}{69.33}{25}{63.67}{69.33}{26}
\emline{63.67}{69.33}{27}{85.00}{76.33}{28}
\special{em:linewidth 0.4pt}
\unitlength 1.00mm
\linethickness{0.4pt}
\put(63.33,96.33){\vector(1,3){0.2}}
\emline{56.33}{76.00}{29}{63.33}{96.33}{30}
\put(63.33,38.67){\vector(1,-4){0.2}}
\emline{56.67}{76.67}{31}{63.33}{38.67}{32}
\put(56.33,127.00){\vector(0,1){0.2}}
\emline{56.33}{76.67}{33}{56.33}{127.00}{34}
\put(16.33,63.33){\vector(-3,-1){0.2}}
\emline{56.33}{77.00}{35}{16.33}{63.33}{36}
\put(52.67,131.00){\makebox(0,0)[cc]{$|c\rangle$}}
\put(10.67,59.00){\makebox(0,0)[cc]{$|a\rangle$}}
\put(106.33,76.67){\makebox(0,0)[cc]{$|b\rangle$}}
\put(67.00,94.34){\makebox(0,0)[cc]{$|\psi_1\rangle$}}
\put(68.00,35.33){\makebox(0,0)[cc]{$|\psi_2\rangle$}}
\unitlength 1.00mm
\linethickness{0.4pt}
\put(68.67,123.00){\makebox(0,0)[cc]{$|\psi_2^{\prime}\rangle$}}
\put(100.33,76.67){\vector(1,0){0.2}}
\emline{57.00}{76.67}{37}{100.33}{76.67}{38}
\put(63.67,117.33){\vector(1,4){0.2}}
\emline{56.67}{76.33}{39}{63.67}{117.33}{40}
\put(21.99,77.33){\makebox(0,0)[cc]{$|\psi_2^{\prime\prime}\rangle$}}
\put(21.67,72.33){\vector(-4,-1){0.2}}
\emline{56.33}{77.00}{41}{21.67}{72.33}{42}
\put(21.67,72.33){\vector(0,1){0.2}}
\emline{21.67}{72.33}{43}{21.67}{72.33}{44}
\put(21.67,72.33){\vector(0,1){0.2}}
\emline{21.67}{72.33}{45}{21.67}{72.33}{46}
\end{picture}

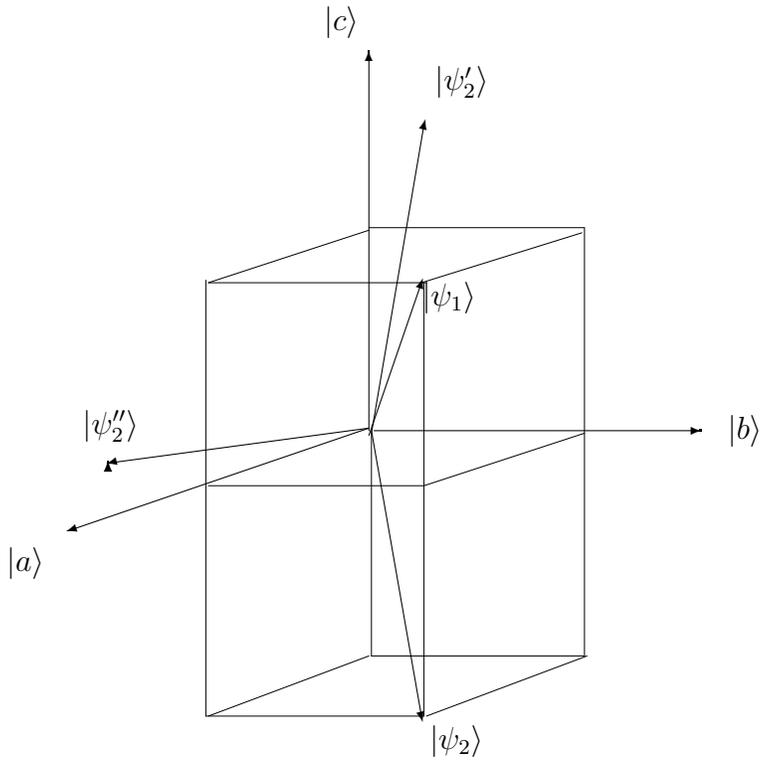
\centerline{Figure 1}
\vskip 2cm

We label the shutters with
 letters a, b, and c. In addition to the shutter location
basis (\{a,b,c\}), the figure also shows
a ``post-selection basis'' $\{\psi_2\}$; i.e., the post-selection state $|\psi_2\rangle$ and two
orthogonal vectors here labeled $|\psi_2^{\prime}\rangle$ and $|\psi_2^{\prime\prime}\rangle$
and defined as follows:

$$|\psi_2^{\prime}\rangle = {1\over \sqrt 6} [|a\rangle
+ |b\rangle + 2|c\rangle]\eqno(2)$$

$$|\psi_2^{\prime\prime}\rangle = {1\over \sqrt 2} [|a\rangle
- |b\rangle \eqno(3)$$

 The shutter particle is pre-selected 
in a superposition of shutter locations, $|\psi_1\rangle = {1\over \sqrt 3} [|a\rangle
+ |b\rangle + |c\rangle]$. The photon is emitted toward the shutters with access
to only 2 of them, in an arbitrary superposition $|\psi_{ph}\rangle = 
[\alpha_1 |a'\rangle + \alpha_2 |b'\rangle]$ (the primed labels
refer to the photon Hilbert space). For simplicity we let
 $\alpha_1 = \alpha_2 = {1\over \sqrt 2}$; the same argument will apply for
arbitrary photon coefficients. 

After the photon has interacted with the shutter particle, the total
combined state can be written as a sum of two terms, one in which 
the photon evaded the shutter particle and one in which it encountered
the shutter particle:

$$|\Psi_{tot}\rangle = {1\over \sqrt 6} \biggl( |a'\rangle |b\rangle +
|a'\rangle |c\rangle + |b'\rangle |a\rangle + |b'\rangle |c\rangle \biggr)
+ {1\over \sqrt 6} \biggl( |a'\rangle |a\rangle +
|b'\rangle |b\rangle \biggr). \eqno (4)$$

In Eq. (4), photon states paired with
the same letter shutter state are reflected. Thus the combined system for
any transmitted photons is characterised by the final state

$$|\Psi_{tr}\rangle = {1\over 2} \biggl(( |a'\rangle |b\rangle +
|a'\rangle |c\rangle + |b'\rangle |a\rangle + |b'\rangle |c\rangle \biggr)
\eqno(5)$$

Heuristically, it is helpful to view this problem as
one in which the three-dimensional shutter Hilbert space, which
corresponds to the standard 3-Box experiment, acquires
a two-fold degeneracy. The degeneracy arises from the entanglement
of the shutter particle with the photon, which is characterised 
by  a 2-dimensional Hilbert space. It corresponds to the fact that
each eigenvalue of shutter particle location now has a sub-index
with two values indicating `a' or `b' for the location of the photon.

Now let us switch from the shutter particle position 
basis $\{a,b,c\}$ to the shutter particle post-selection basis $\{\psi_2\}$
and introduce the two-fold degeneracy due to the photon, so that the eigenspace 
defined by each of the shutter particle basis vectors expands to a plane. Using
the notation
$|\psi_{2a}\rangle = |a'\rangle \otimes |\psi_2\rangle$ and
$|\psi_{2b}\rangle = |b'\rangle \otimes |\psi_2\rangle$, etc.,
the resulting basis vectors for the combined
Hilbert space are as follows:

$$|\psi_{2a}\rangle =\!\left(\matrix{1\cr 0\cr 0\cr 0\cr 0\cr 0 }\right);
|\psi_{2b}\rangle =\!\left(\matrix{0\cr 1\cr 0\cr 0\cr 0\cr 0 }\right);
|\psi_{2a}^{\prime}\rangle =\!\left(\matrix{0\cr 0\cr 1\cr 0\cr 0\cr 0 }\right);
|\psi_{2b}^{\prime}\rangle =\!\left(\matrix{0\cr 0\cr 0\cr 1\cr 0\cr 0 }\right);
|\psi_{2a}^{\prime\prime}\rangle =\!\left(\matrix{0\cr 0\cr 0\cr 0\cr 1\cr 0 }\right);
|\psi_{2b}^{\prime\prime}\rangle =\!\left(\matrix{0\cr 0\cr 0\cr 0\cr 0\cr 1 }\right)\eqno(6)$$

\noindent 
and the post-selection shutter subspace $|\psi_2\rangle$
becomes a 2-dimensional subspace spanned by the two vectors
$|\psi_{2a}\rangle$ and $|\psi_{2b}\rangle$,
which is orthogonal to the final
state $|\Psi_{tr}\rangle$ of the system for transmitted 
photons. This can readily be seen by 
rewriting $|\Psi_{tr}\rangle$
in terms of the shutter post-selection basis $\{\Psi_2\}$ in the total
 (6-dimensional) Hilbert space:

$$|\Psi_{tr}\rangle_{\{\psi_2\}} = \left(\matrix{0\cr 0\cr {\sqrt 6\over 4}\cr
{\sqrt 6\over 4}\cr {-\sqrt 2\over 4}\cr {\sqrt 2\over 4}}\right)\eqno(7)$$

Since the final state corresponding to photon transmission is orthogonal
to the subspace corresponding to post-selection, 
the same purpose is served as in the
standard 3-Box experiment: to throw out runs of the experiment in which the
particle's final state does not support the desired property (in this case,
reflection of the photon).
Any shutter particle which is successfully post-selected 
corresponds to a photon which could not have been transmitted and therefore must have 
been reflected. Thus, successful post-selection in an experiment
of this kind (that is, using both the shutter particle and the photon)
 means no more than the following: `on {\it this particular run}, 
the photon and the shutter particle 
encountered each other in a standard quantum superposition
of slits and the photon was reflected'. There is no basis
 for the assertion that the proposed experiment 
``makes the claim that such a particle is simultaneously
in several places even more robust,'' (Aharonov and
Vaidman (2002, p.1), since it is
not the case here that the shutter particle is ``with certainty''
in each of two different slits. 

 One can discern here two distinct ways in which
Aharonov and Vaidman apply the term ``with certainty'': the first,
that the shutter particle
reflects the photon ``with certainty,'' and the second,
that the shutter particle can (somehow, paradoxically) 
be thought of as being in each of both slits $a$ and $b$
with certainty.  Regarding the first usage, their
 claim that the ``pre- and post-selected shutter reflects the photon with
 certainty'' is tautological, because the post-selection measurement
essentially functions as a measurement
of whether the photon was reflected. (Of course,
not all shutter particles corresponding to reflected photons will be post-selected, just
as in the three-box case where not all particles found in the opened box
will be post-selected;\footnotemark[4]
post-selection of the shutter particle is a sufficient but not necessary
criterion for photon reflection.)
Thus, the claim of ``certainty'' in this context has no more content than ascertaining
a certain measurement result---i.e., that the
photon was reflected---and, after the fact, claiming that
the photon was reflected ``with certainty.'' But of course all measurement
outcomes are certain once we have obtained them. 

As for the second usage of ``with certainty'': the fact that
the photon and the shutter particle can be in the same superposition of shutter states
adds nothing beyond what we already know about quantum superpositions,
such as the strange properties of the two-slit experiment. 
That is, the only valid sense in which the shutter particle can
be said to occupy ``both shutters'' is the same one
in which a particle can be said to have gone through
``both slits'' in the standard two-slit experiment. 

Aharonov and Vaidman perhaps want to invoke the idea that time
symmetry of pre- and post-selection implies that a shutter particle
successfully post-selected at time $t_2$ was ``fated'' to be post-selected even at 
time $t < t_2$. However, adding this assumption implies only
that such a shutter particle was fated to encounter the photon
(or to be in the same superposition of shutters as the photon), not the
stronger conclusion that the particle occupies ``all N shutters at once.''
 
Aharonov and Vaidman's presentation conflates an interesting 
and valid result---the correlation of the photon and the shutter particle in an 
identical superposition of slits---with a questionable previous
claim about the ``certainty'' of finding a particle in
two different locations if it were possible to look for 
the same pre- and post-selected particle separately in each. Though
Aharonov and Vaidman say that the photon is performing a
``strong'' measurement,\footnotemark[5] this is misleading because
the photon, being in a superposition, is not measuring the shutter particle's 
definite presence in any particular slit. That is, the photon is
measuring neither observable A (corresponding to looking for
the shutter particle only in slit
$a$) nor observable B (corresponding to looking only in
slit $b$). The photon therefore cannot act
as a surrogate for the experimenter who would like to test
the surprising claim of the particle's simultaneous definite presence
in those two different locations. Thus the present experiment, while 
interesting, does not constitute evidence for robustness 
of the latter claim.
\vskip 1cm
Acknowledgements.

The author gratefully acknowledges valuable correspondence with Jeffrey Bub
and Jerry Finkelstein, and helpful criticism from several anonymous reviewers
(any remaining flaws in this paper are, of course, solely the author's responsibility).

This work was supported in part by NSF Grant no. SES-0115185.
\newpage\onehalfspacing
\noindent $^1$The Born Rule tells us
that the probability for outcome $q_k$ when observable Q is
measured on a system prepared in state $|\psi\rangle$
is given by $|\langle q_k|\psi\rangle|^2$.

\noindent $^2$Aharonov and Vaidman (2002, p.3).

\noindent $^3$Of course, adhering strictly to the name ``N-box experiment'' would
require us to call this the ``2-shutter experiment.''

\noindent $^4$The ``reduced'' density operator
corresponding to photon reflection in the Hilbert space of the shutter particle is
$W_{r, sh} = Tr_{ph} [{1\over 2} (|a'\rangle |a\rangle \langle a|\langle a'| +
 |a'\rangle |a\rangle \langle b|
\langle b'| + |b'\rangle |b\rangle \langle a|\langle a'| + |b'\rangle |b\rangle \langle b
|\langle b'|)]
= {1\over 2}(|a\rangle \langle a| + |b\rangle \langle b|)$; the probability that
a shutter particle in this state will be post-selected is
$Tr[|\psi_2\rangle\langle\psi_2|W_{r,sh}] = {1\over 3}.$

\noindent $^5$As opposed to 
a ``weak measurement'' in which the interaction Hamiltonian
between the system and measuring device is weakened; cf.
Aharonov and Vaidman (1990).

\newpage
References

\noindent Aharonov, Y. and Vaidman, L. (2002)
How One Shutter Can Close N Slits, e-print: www.arxiv.org,
quant-ph/0206074. \newline
Aharonov, Y. and Vaidman, L. (1990) Properties of a quantum system during
the time interval between two measurements, {\it Physical Review A}, 41,
pp. 11-20. \newline
Aharonov, Y. and Vaidman, L. (1991) Complete Description
of a Quantum System at a Given Time,
{\it Journal of Physics A}, 24, pp.2315-2328.\newline
Cohen, O. (1995) Pre- and postselected quantum systems, 
counterfactual measurements, and consistent histories, 
{\it Physical Review A}, 51, pp. 4373-4380.\newline
Griffiths, R. B. (1996) Consistent Histories and Quantum Reasoning,
{\it Phys. Rev. A}, 54, pp. 2759-2774.\newline
Kastner, R. E. (1999a) Time-Symmetrised Quantum Theory, Counterfactuals
and Advanced Action, {\it Studies in History and Philosophy of Modern Physics},
30, pp. 237-259.\newline
Kastner, R. E. (1999b) The Three-Box Paradox and Other Reasons to Reject
the Counterfactual Usage of the ABL Rule, {\it Foundations of Physics},
29, pp. 851-863.\newline
Vaidman, L. (1999a) The Meaning of Elements of 
Reality and Quantum Counterfactuals -- Reply to Kastner,
{\it Foundations of Physics}, 29, pp. 865-876.\newline
Vaidman, L. (1999b) Defending Time-Symmetrized Quantum Counterfactuals,
{\it Studies in History and Philosophy of Modern Physics}, 30, pp. 373-397.

\newpage

Biographical Note

Ruth E. Kastner is a Research Associate in the Department of
Philosophy at the University of 
Maryland at College Park. The present work was partly funded by
a grant from the Science and Technology Studies 
Division of the National Science Foundation in 2001.
She is continuing her study of time symmetry in quantum mechanics,
focusing in particular on John Cramer's Transactional Interpretation,
under a second NSF grant. 
\vskip .2cm\noindent 
Department of Philosophy\\
University of Maryland\\
College Park, MD 20742 USA\\
rkastner@wam.umd.edu

\end{document}